\documentclass[conference]{IEEEtran}
\IEEEoverridecommandlockouts
\ifCLASSOPTIONcompsoc
\usepackage[caption=false,font=normalsize,labelfon
t=sf,textfont=sf]{subfig}
\else
\usepackage[caption=false,font=footnotesize]{subfig}
\fi
\usepackage{cite}
\usepackage{amsmath,amssymb,amsfonts}
\usepackage{algorithmic}
\usepackage{graphicx}
\usepackage{xcolor}
\usepackage{textcomp}
\usepackage{xcolor}
\usepackage{tabularx}
\usepackage{multirow}
\usepackage{float}
\usepackage{caption}
\usepackage{lipsum}
\usepackage{physics}
\def\BibTeX{{\rm B\kern-.05em{\sc i\kern-.025em b}\kern-.08em
    T\kern-.1667em\lower.7ex\hbox{E}\kern-.125emX}}
\begin{document}

\title{Efficient VQE Approach for Accurate Simulations on the Kagome Lattice \\
}


\makeatletter
\def\endthebibliography{%
  \def\@noitemerr{\@latex@warning{Empty `thebibliography' environment}}%
  \endlist
}
\makeatother

\author{
\IEEEauthorblockN{Jyothikamalesh S}
\IEEEauthorblockA{\textit{Sri Eshwar College of Engineering} \\
Tamil Nadu, India \\
jyothikamaleshs@gmail.com}
\and
\IEEEauthorblockN{Kaarnika A}
\IEEEauthorblockA{\textit{Sri Eshwar College of Engineering} \\
Tamil Nadu, India \\
kaarnikaa2003@gmail.com}
\and
\IEEEauthorblockN{Dr.Mohankumar.M}
\IEEEauthorblockA{\textit{Sri Eshwar College of Engineering} \\
Tamil Nadu, India \\
mail2mohanphd@gmail.com}
\and
\IEEEauthorblockN{Sanjay Vishwakarma}
\IEEEauthorblockA{\textit{IBM Quantum} \\
California, USA \\
sanjay.vishwakarma@ibm.com}
\and
\IEEEauthorblockN{Srinjoy Ganguly}
\IEEEauthorblockA{\textit{Woxsen University} \\
Hyderabad, India \\
srinjoy.ganguly@woxsen.edu.in}
\and
\IEEEauthorblockN{Yuvaraj P}
\IEEEauthorblockA{\textit{Independent Researcher} \\
Tamil Nadu, India \\
pyuvaraj1372000@gmail.com}
}

\maketitle

\begin{abstract}
The Kagome lattice, a captivating lattice structure composed of interconnected triangles with frustrated magnetic properties, has garnered considerable interest in condensed matter physics, quantum magnetism, and quantum computing.The Ansatz optimization provided in this study along with extensive research on optimisation technique results us with high accuracy. This study focuses on using multiple ansatz models to create an effective Variational Quantum Eigensolver (VQE) on the Kagome lattice. By comparing various optimisation methods and optimising the VQE ansatz models, the main goal is to estimate ground state attributes with high accuracy. This study advances quantum computing and advances our knowledge of quantum materials with complex lattice structures by taking advantage of the distinctive geometric configuration and features of the Kagome lattice. Aiming to improve the effectiveness and accuracy of VQE implementations, the study examines how Ansatz Modelling, quantum effects, and optimization techniques interact in VQE algorithm. The findings and understandings from this study provide useful direction for upcoming improvements in quantum algorithms,quantum machine learning and the investigation of quantum materials on the Kagome Lattice.
\end{abstract}


\section{Introduction}
The Kagome lattice, characterized by its captivating interplay of interconnected triangular units, has captured the attention of researchers and enthusiasts worldwide. This visually striking lattice configuration possesses remarkable geometric and electronic attributes, rendering it a captivating object of study across diverse disciplines such as physics, material science, and quantum computing. Its intricate nature continues to spark innovative breakthroughs and propel scientific progress.

Moreover, the Kagome lattice serves as a fascinating testbed for quantum algorithms like the Variational Quantum Eigensolver (VQE). By leveraging the unique geometry and properties of the lattice, VQE offers a promising avenue for exploring the elusive ground state and energy landscape of the Kagome system. Pursuing VQE on the Kagome lattice not only unlocks novel insights into fundamental physics but also holds the potential to revolutionize materials design, quantum simulations, and information processing. Embracing this research frontier empowers scientists to unravel the mysteries of the Kagome lattice and unleash its immense technological potential.

Due to geometric frustration, the Kagome lattice, which consists of a network of triangles, displays remarkable physical phenomena. It has become a fascinating platform for researching topological states, quantum spin liquids, and other quantum processes. Using a hybrid evolutionary algorithm (HEA), we implement VQE on the Kagome lattice in order to investigate its quantum features and offer insights into its behaviour.

The HEA can effectively explore the VQE ansatz's parameter space  by combining the power of classical optimisation algorithms with quantum computations. This method may be able to get around the difficulties in determining the ansatz's optimal parameters, improving the accuracy of estimates of the ground state properties. With the primary goal of achieving high accuracy in estimating ground state energies and characteristics, we use HEA to improve the VQE implementation on the Kagome lattice in this study.

We examine a variety of optimizers, such as BFGS, SPSA, and Cobyla, which are frequently used in VQE implementations,to assess the efficacy and performance of our technique. Each optimizer has unique advantages and disadvantages in terms of resilience, robustness to noise, and capacity to manage local minima. We compare their Kagome lattice performance in order to determine which optimizer provides the most accurate results with the fastest convergence.

We examine several alternative configurations for the VQE implementation on the Kagome lattice in addition to optimising the optimizer selection. We contrast the HEA-generated ansatz with the widely utilised EfficientSU2 and UCCSD ansatz. While UCCSD offers realistic descriptions of electronic systems, EfficientSU2 is renowned for its efficiency and ability to capture quantum correlations. By adding HEA, we seek to increase the ansatz's expressibility and especially tailor it to the unique properties of the Kagome lattice.

Through this study, we aim to show that a VQE implementation that is effective and uses HEA can produce estimates of ground state properties on the Kagome lattice that are highly accurate. We compare several optimizers and ansatz configurations in order to find the ideal mix that, while taking computing efficiency into account, yields the best accuracy. This study's findings can aid in the comprehension of quantum materials, quantum simulations, and the creation of more potent quantum algorithms.

In the sections that follow, we will outline the procedures and methods we utilised, show the findings of our tests, and talk about the consequences and future directions of our research. We aim to improve both our understanding of quantum materials with complex lattice structures as well as the field of quantum computing in general by examining the VQE implementation on the Kagome lattice using HEA.

\section{Kagome Lattice}
An intriguing lattice structure with a distinctive configuration of interconnecting triangles is the Kagome lattice. It has received abundant attention from researchers in many different domains, including condensed matter physics, quantum materials, and quantum computing.

Condensed matter Physics researchers are very interested in the mineral complex known as Herbertsmithite. It is made up of copper, zinc, and hydroxide ions and is a natural manifestation of the Kagome lattice structure. $Cu_3Zn(OH)_6Cl_2$ is the chemical formula for it.

\begin{figure}[!t]
\centering
\includegraphics[width=3.5in]{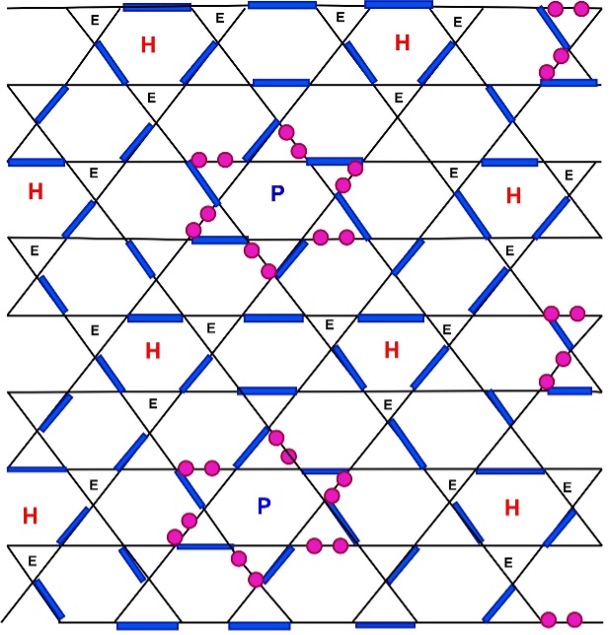}
\caption{The kagome-lattice Heisenberg model displays a distinctive pattern of bonds, with the bonds associated with the lowest energy levels represented by blue or dark grey colors, corresponding to the fundamental state energy.Pinwheels, empty triangles, and perfect hexagons are denoted by P, E, and H respectively. The presence of dark solid blue or dark grey bonds, as well as magenta/grey dotted bonds, indicates the two dimer coverings of pinwheels, which can exhibit high-order degeneracy in perturbation theory.}
\label{kagome-lattice}
\end{figure}

The Heisenberg model H, on a kagome lattice, featuring spin-dependent couplings among spin-1/2 particles.Various methods have been employed to study its properties, both numerically and analytically. The KLHM exhibits intricate interactions between spins, leading to interesting ground state properties. 
\begin{equation*}
    H = J \sum_{<i,j>} S_i*S_j  
\end{equation*} 

\cite{PhysRevB.106.214429}
quantum spin model has been a source of significant frustration, as researchers have encountered challenges in fully understanding its characteristics, particularly the nature of its ground state. Despite extensive computational and theoretical approaches utilized to investigate the model, the exact properties of its ground state continue to be a subject of discussion. Various possibilities, including valence bond crystals (VBC) and spin-liquid states with algebraic correlations, have been suggested. Recent experimental investigations on the compound $Cu_3Zn(OH)_6Cl_2$ have added to the intrigue surrounding this model.\cite{PhysRevB.76.180407}.

Herbertsmithite has remarkable magnetic properties as a result of the Kagome lattice's distinctive geometric configuration. The interactions between neighbouring magnetic moments cannot be satisfied at the same time due to a phenomenon known as geometric frustration, leading to a highly disordered ground state. This dissatisfaction leads to a spin-liquid state by preventing the development of a conventional magnetic order.

Long-range magnetic moment entanglement and quantum fluctuations characterise the spin liquid state that has been found in Herbertsmithite. Herbertsmithite is a perfect solution for researching quantum spin liquids, states of matter where quantum entanglement is a defining characteristic.
The kagome lattice was designed with the IBM Gualape 16-qubit hardware in mind, although this mapping is expandable to greater numbers of qubits.

\begin{figure}[!t]
\centering
\includegraphics[height=2.7in,width=3.2in]{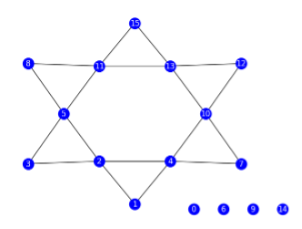}
\caption{Hardware realization of Kagome unit cell in a 16 qubit hardware}
\label{unit-cell}
\end{figure}

\begin{figure}[!t]
\centering
\includegraphics[height=2.8in,width=3.2in]{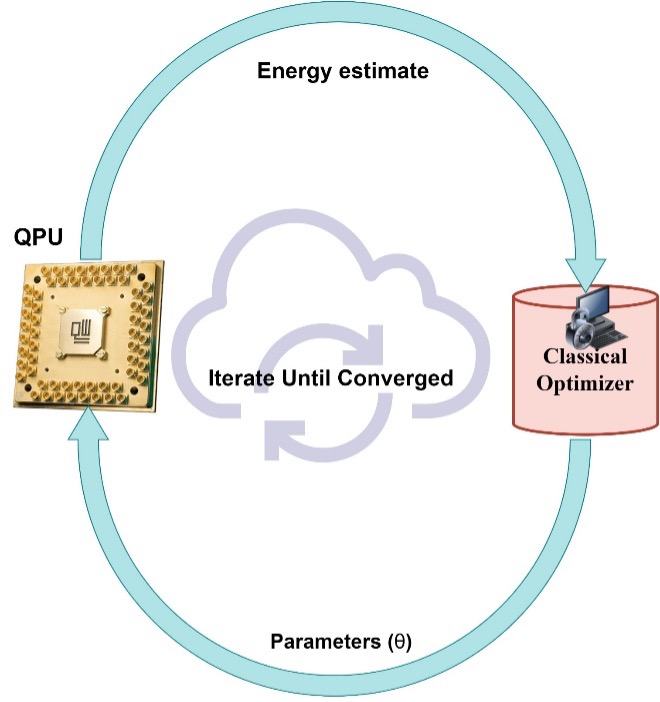}
\caption{VQE algorithm minimizes the energy represented by $E_0(\theta)$ by finding the optimal values for the parameters $\theta$. The classical optimization algorithm progressively fine-tunes these parameters while utilizing a quantum chip to calculate the expected value of the Hamiltonian H. The objective function is specifically designed to assess the anticipated value of the simulated Hamiltonian H. }
\label{algorithm_schematic}
\end{figure}

\cite{lavrijsen_classical_2020}. 

\section{Quantum variational eigensolver}
The Variational Quantum Eigensolver (VQE) is an intriguing hybrid quantum algorithm that synergistically harnesses the computational power of quantum computers along with  optimization techniques pertaining to classical computation.
At the heart of the VQE algorithm resides an ansatz, which is a parameterized quantum circuit representing a trial wavefunction. By carefully adjusting the parameters of the ansatz, the algorithm seeks to minimize the system's energy. Classical optimization methods are commonly employed to optimize these parameters. In each iteration, the quantum circuit is executed to measure the expectation value of the Hamiltonian, providing an approximation of the ground-state energy. The classical optimizer then updates the ansatz parameters based on these measurements, iteratively refining the estimation. Through the harmonious interplay between quantum computation and classical optimization, the VQE algorithm aspires to attain a precise estimation of the ground-state energy for the given Hamiltonian, paving the way for diverse applications in quantum computing and beyond.

\begin{figure}[!t]
\centering
\includegraphics[height=2.8in,width=3.2in]{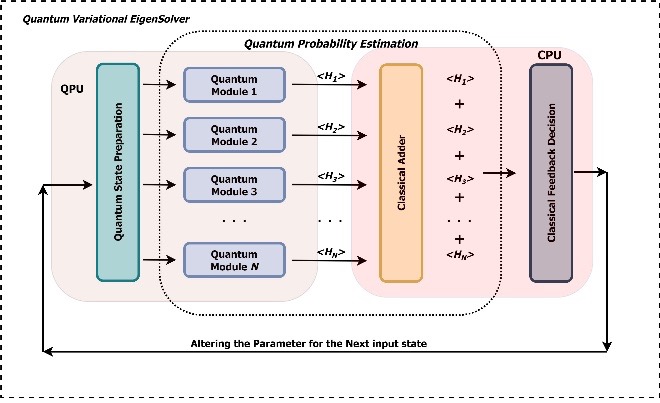}
\caption{VQE algorithm schematic}
\label{Architecture of the quantum-variational eigensolver}
\end{figure}

Quantum modules in QEE use previously produced quantum states to compute $\langle H_i \rangle$, where $\langle H_i \rangle$ is any individual term in the sum defining H. The CPU computes using the results after passing them along. The classical minimization process, which is run on the CPU, uses the quantum variational eigensolver To determine the updated state parameters, which are subsequently transmitted back to the quantum processing unit (QPU), 

\section{Hamiltonian}
The first quantized form of the Hamiltonian can therefore be written directly in the single particle basis \cite{tilly_variational_2022}.
    
\begin{align*}
   \Hat{H} &= \sum_{i=1}^{m} \sum_{p,q=1}^{n} h_{p,q} |\phi_p^i \rangle \langle \phi_q^i| \\ &+ \frac{1}{2} \sum_{i\neq j}^{m}\sum_{p,q,r,s=1}^{n} h_{pqrs} |\phi_p^i \phi_q^j \rangle \langle \phi_r^i \phi_s^j |
\end{align*}

The Heisenberg model, a fundamental concept in quantum physics, elucidates the intricate interactions among quantum spins in a lattice. The captivating Kagome lattice, characterized by interwoven triangles forming a hexagonal structure, provides a captivating backdrop for exploring the Heisenberg model's dynamics.

To capture the essence of the Heisenberg model on the Kagome lattice, the Qiskit package offers the Ising class—a powerful framework for simulating spin systems. Leveraging the Ising class, researchers can precisely define the exchange interactions between spins on the Kagome lattice, incorporating the requisite terms of the Hamiltonian.

Through the utilization of the Ising class in Qiskit, one can create a faithful representation of the Heisenberg model on the Kagome lattice, accounting for both the spin states and their associated angular momentum. This empowers investigations into captivating phenomena like spin correlations, magnetic ordering, and quantum phase transitions unique to the Kagome lattice.

By conducting simulations utilizing the Ising class, researchers can delve into the intricate nuances of ground state properties, energy spectra, and dynamic behavior within the Heisenberg model on the Kagome lattice. These explorations foster deeper insights into quantum magnetism, condensed matter physics, and pave the way for groundbreaking quantum algorithms tailored for the Kagome lattice.

In essence, the Ising class in Qiskit stands as an invaluable resource, enabling the creation and simulation of the Heisenberg model on the captivating Kagome lattice. This sophisticated framework empowers researchers to unravel the profound mysteries of spin systems, ushering in new horizons for quantum computing and materials science.

\subsubsection{Lattice Hamiltonians}

Lattice Hamiltonians: Lattice models describes a physical system's behaviour that arises from the interaction's within the lattice structure rather than constructing a Hamiltonian to mathematically represent with specific parameters . The "particles moving in this discretized space" in this case are electrons. It should be noted that representation and encoding would be significantly simpler if bosonic particles were taken into account as opposed to fermionic particles because they do not require the fermionic antisymmetric interactions that are discussed in the following section. Condensed matter physics frequently employs lattice models to simulate the phenomenological characteristics of particular materials, including phase transitions and electronic band structures. There are many other lattice models; we simply briefly discuss a handful here:

2) Spin Hamiltonians, such as the Heisenberg model 

\begin{equation*}
    \Hat{H} = J \sum_{\langle p,q \rangle} \hat{S}_p, \hat{S}_q
\end{equation*}

In the given expression, the term $\sum_{\langle p, q \rangle}$ represents a sum over neighboring pairs of sites on the lattice. The variable J is a constant denoting , and $\hat{S}_p=(\hat{S}_p^x, \hat{S}_p^y, \hat{S}_p^z)$ represents the three spin-1/2 angular momentum operators on site p. It is important to note that the spin-1/2 matrices are related to the Pauli matrices through the equation $(\hat{S}^x,\hat{S}^y, \hat{S}^z )= \frac{h}{2}(X,Y,Z)$. It is worth emphasizing that these mathematical expressions capture the fundamental interactions and dynamics of the system under consideration. \cite{tilly_variational_2022}

\section{Ansatz}

The ansatz assumes a critical role within the VQE framework as it provides a parameterized quantum circuit, serving as a trial wavefunction. This ansatz encompasses a variational manifold of quantum states, adeptly prepared on a quantum computing platform.

In the context of the VQE algorithm, the ansatz serves as an initial configuration for the optimization process, aimed at identifying the optimal set of parameters that yield minimized energy for the quantum system under scrutiny. Through parameter adjustments within the ansatz circuit, the algorithm ventures into the intricate realm of quantum states, paving the way for exploration of diverse potential solutions.

The selection of an appropriate ansatz entails profound significance, as it determines the expressiveness and complexity of the resulting quantum circuit. A well-crafted ansatz strikes an intricate balance between flexibility and computational efficiency, effectively capturing the salient attributes of the system's ground state while simultaneously managing the circuit's size to ensure tractability and feasibility.

When implementing the unitary coupled cluster ansatz on a quantum computer, it is crucial to address the effects of discretization errors inherent in the approximation methods used.. \cite{kandala_hardware-efficient_2017} This gives us insight into error mitigation approaches that we have to consider for UCCSD ansatz 

One active research area towards this objective to is the study of the expressive power of different ansatz constructions because an ansatz with a strong expressive power can represent more complicated functions \cite{wu_towards_2021}, We developed an ansatz architecture specifying to kagome lattice with hardware efficiency as the focus which can perform complex functions and represent all The possible outcomes in Hilbert space better

Considering  a quantum mechanical system characterized by a Hilbert space denoted as $H$ with a dimensionality of $N$. The system is governed by a Hamiltonian, denoted as $H$, which possesses a ground state energy represented by $E_0$. Within this context, we examine a specific subset of states denoted as ${|\theta \rangle }$, where the states are parameterized by $\theta$ belonging to the real-valued space $R_m$. It is crucial to note that for a comprehensive representation of the entire Hilbert space $H$, the parameter space $m$ must scale proportionally to $N$, denoted as $O(N)$.

\begin{equation*}
    E(\theta) = \langle \theta| H |\theta \rangle \geq E_0 
\end{equation*}

is the variational principle, which embraces the entirety of $|\theta \rangle$ states within the expansive Hilbert space $H$, stands as a fundamental tenet in the realm of quantum mechanics. In pursuit of scalability, variational methods judiciously employ a set of states endowed with $m = polylog N$ parameters. Within this framework, prominent methodologies such as the VQE endeavor to meticulously minimize the intricate energy function $E(\theta)$, Consequently, revealing a priceless upper limit on the enigmatic ground-state energy..\cite{PhysRevB.106.214429}

\subsection{Hardware Effecient Ansatz}
The Hardware-Efficient Ansatz (HEA) emerged from the objective of parameterizing the trial state in VQE using custom-designed quantum gates tailored to the specific quantum hardware. Various iterations of the Hybrid Entangled Ansatz (HEA) have been put forward, sharing a common strategy of constructing the ansatz by interlinking blocks of individual-qubit operations, adjustable parameter-based rotation gates, and entangling gates. 
The selection of specific rotation and entangling gates is contingent upon the native gate set available on the quantum device being utilized, as well as the intended intricacy of the target quantum state. HEA offers the advantage of being expressive while adapting to the device's native gate set, making it widely used in small-scale quantum research.

However, HEA has limitations. Creating an indicative ground state wavefunction necessitates traversing a substantial portion of the Hilbert space., which can be inefficient. In some cases, achieving sufficient accuracy may require an exponential depth. Further research is needed to assess the accuracy of the ground value under such conditions.

The HEA for the kagome lattice structure is predominated by Rotational gated to represent the hilbert space and almost reduced the usage of entangling states that we can say it's influence upon the algorithm is naught.

\subsection{UCCSD -The Unitary Coupled Cluster Singles and Doubles  ansatz}

The Unitary Coupled Cluster Singles and Doubles (UCCSD) ansatz serves as a parameterized trial state in VQE, utilizing unitary transformations to encompass the intricate quantum correlations among electrons. UCCSD constructs the trial state by incorporating single and double excitations, reflecting the electron movement between distinct molecular orbitals. Renowned for its expressive power and aptitude in capturing electronic correlations, the UCCSD ansatz holds a prominent position in quantum chemistry applications. Nonetheless, the UCCSD ansatz encounters challenges stemming from the exponential growth of the Hilbert space, intimately tied to the number of electrons and orbitals involved. Extensive investigation is warranted to gauge the precision of ground-state energy estimation under such conditions.

\subsection{Efficient SU2 ansatz}

The EfficientSU2 ansatz epitomizes a remarkable approach for parameterizing the trial state in VQE, characterized by its exceptional capability in quantum state preparation. This ansatz leverages a well-designed parameterized quantum circuit structure, seamlessly incorporating a concise ensemble of gates and entangling operations, meticulously tailored to optimize computational efficiency while preserving expressive power. The EfficientSU2 ansatz, renowned for its versatility and adaptability across diverse quantum hardware architectures, boasts the potential to yield compact circuit implementations. This remarkable feature renders it an invaluable contender for VQE applications, meriting meticulous investigation to comprehend its nuanced performance characteristics and ascertain its suitability for specific quantum systems.

\begin{table*}[tbh!]
\caption{Comparison of various ansatz and their depth, parameter and entangling gate complexity to how well they performed for the Kagome lattice specifications with different types of optimizers.}
\label{tab:LSTMVariants}
\begin{tabular}{p{.15\textwidth}p{.15\textwidth}p{.2\textwidth}p{.2\textwidth}p{.2\textwidth}}
\textbf{Method} & \textbf{Depth} & \textbf{Parameters} & \textbf{Entangling gates}  & \textbf{Comments} \\\hline
Hardware Efficient Ansatz (HEA) & $O(x)$ & $O(Nx)$ & $O((N-1)x)$   & x is the layers of gates required to represent the Hilbert space and in the worst case we will have to represent the whole of hilbert space to find the fundamental energy level

\\ UCCSD & $O((N-m)^2 m\tau)$ & $O((N-m)^2 m^2 \tau)$  & $O(2(\Tilde{q}-1)^2 N^2 \tau)$ & $q$ is the average of the Pauli weights involved in constructing the ansatz.  $N$ Symbolises the maxima of Pauli weight under Jordan-wigner mapping, 
and $log(N)$ is below Bravyi-Kitaev. $\tau$ denotes the iterations involved in Trotterization process

\\ EffecientSU2 & $O(N)$ & $O(N)$ & $O(N)$ & Effecientsue is directly proportional to the no of qubits and operations performed with it.EffecientSU2 ansatz is available in the Qiskit package and EfficientSU2 ansatz is the realization of Hardware focused ansatz  available in Qiskit package \\

\end{tabular}
\end{table*}

\section{Optimiser}
The optimizer assumes a vital role in VQE, adjusting circuit parameters to minimize the objective function and accurately estimate the Fundamental energy level. The objective function represents the disparity between measured expectation values and the true ground state energy . We compared BFGS, COBYLA, and SPSA optimization techniques for comprehensive research.

\subsection{Constrained Optimisation by Linear (COBYLA)}

\begin{figure}[!t]
\centering
\includegraphics[width=3.5in]{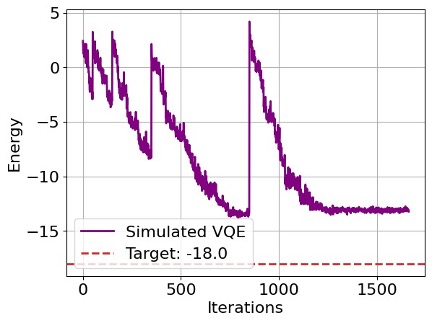}
\caption{Energy levels in Hartree plotted with respect to iterations in Y and X axis respectively for COBYLA optimiser}
\label{kagome-lattice}
\end{figure}

 COBYLA, a constrained nonlinear optimization algorithm, iteratively updates the solution estimate using a trust region technique. It solves quadratic programming subproblems, adjusts the solution estimate, and modifies the trust region radius to handle constraints. When combined with a hardware-efficient Ansatz in the context of VQE, COBYLA achieves an accuracy of 80

\subsection{Broyden-Fletcher-Goldfarb-Shanno (BFGS)}

\begin{figure}[!t]
\centering
\includegraphics[width=3.5in]{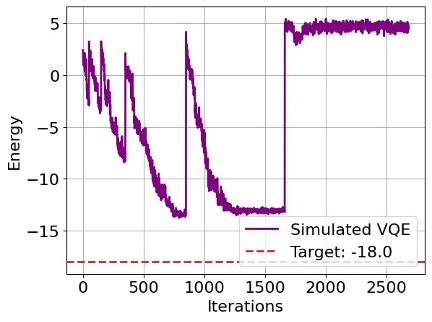}
\caption{Energy levels in Hartree plotted with respect to iterations in Y and X axis respectively for BFGS optimiser }
\label{kagome-lattice}
\end{figure}

For the purpose of addressing unrestricted nonlinear optimisation problems, the BFGS (Broyden-Fletcher-Goldfarb-Shanno) optimizer employs an iterative process. It is a member of the family of line search methods, which is a subset of the quasi-Newton methods. The BFGS optimizer uses the differences between subsequent gradient vectors to roughly approximate the cost function denoting the Hessian matrix.

The BFGS optimizer seeks to discover the accurate solution to an unconstrained nonlinear optimisation iteratively by updating the solution estimate, search direction, step size, and approximate Hessian matrix. In large-scale situations where computing the precise Hessian matrix is expensive or impracticable, the technique updates the Hessian approximation using data from the gradients and solution steps.

\subsection{Simultaneous perturbation stochastic approximation (SPSA)}

\begin{figure}[!t]
\centering
\includegraphics[width=3.5in]{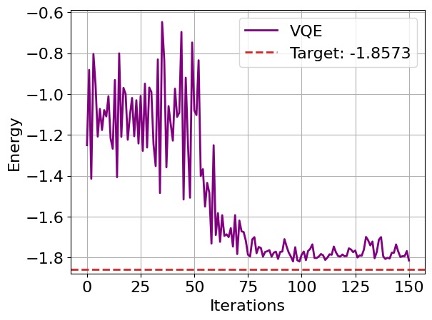}
\caption{Energy levels in Hartree plotted with respect to iterations in Y and X axis respectively for SPSA optimiser}
\label{kagome-lattice}
\end{figure}

The SPSA algorithm is purposefully developed to estimate the gradient of an objective function using a limited number of measurements, precisely two. Its design allows for an efficient approximation of the gradient without the need for extensive data collection. Instead of employing symmetrically chosen measurement points, SPSA introduces a small random vector as a perturbation to the objective function. This unique approach enables SPSA to efficiently estimate the gradient using a limited number of measurements. By incorporating randomness into the perturbation process, the algorithm explores the objective function in a stochastic manner, making it well-suited for diverse optimization tasks. This characteristic allows SPSA to handle various types of optimization problems effectively while maintaining a concise and resource-efficient implementation.

\begin{equation*}
(g(\theta_t))_j = \frac{\mathcal{L}(\theta + c_t \Delta_j) - \mathcal{L}(\theta - c_t \Delta_j)}{2 c_t (\Delta_t)_j}
\end{equation*} 

  The stochastic perturbation vector, denoted as $\Delta t$, assumes a pivotal role in the SPSA algorithm. This algorithm, recognized for its effectiveness in quantum variational models, distinguishes itself by relying on a minimal pair of measurement points per iteration. .SPSA can also be utilised in Noisy optimisation
SPSA approximation with hardware-efficient Ansatz usage in VQE results us in 99 percent accuracy.

The approximation of second-order optimization for optimised gradient is possible with SPSA .

\section{Conclusion}

In the pursuit of constructing an efficient Variaional Quantum Eigensolver(VQE) for the Kagome lattice, various optimizers and Ansatz were compared to assess their performance. The optimizers evaluated included BFGS, COBYLA, and SPSA and the ansatz utilised in the research are Hardware effecient ansatz(HEA),UCCSD ansatz,BFGS ansatz.Through rigorous experimentation and analysis, the research showcased the strengths and weaknesses of each optimizer in terms of convergence speed, accuracy, and robustness. This comprehensive comparison allowed for informed decision-making in selecting the most suitable optimizer for the specific requirements of the Kagome lattice problem. By leveraging the power of advanced optimization techniques, the research has propelled the development of highly effective approaches for tackling complex quantum systems.

\bibliographystyle{IEEEtran}
\bibliography{References}

\end{document}